\documentclass[fleqn,11pt,letterpaper]{article}
\usepackage{graphicx}
\usepackage{amscd}
\usepackage{amsmath,amssymb,amsthm,amsfonts}
\usepackage{bbm}
\usepackage{verbatim}
\usepackage{varioref}
\usepackage{bm}

\usepackage{hyperref}
\begin{document}
	\title{A Second Look at Post Crisis Pricing of Derivatives - Part I: \\
	A Note on Money Accounts and Collateral}
	\author{\Large Hovik Tumasyan\\	
	FinRisk Solutions\footnote{Contact email: htumasyan@finrisksolutions.ca} \footnote{The author would like to thank John Hull, Alan White, Andrew Green and Darrell Duffie for stimulating discussions at different stages of this work. The paper represents personal views of the author and does not constitute an advice.}\\
	}

	\maketitle
	\thispagestyle{empty}
\begin{abstract}
The paper reviews origins of the approach to pricing derivatives post-crisis by following three papers that have received wide acceptance from practitioners as the theoretical foundations for it - \cite{pit}, \cite{BK} and \cite{BK13}. \\
\indent The review reveals several conceptual and technical inconsistencies with the approaches taken in these papers. In particular, a key component of the approach - prescription of cost components to a risk-free money account, generates derivative prices that are not cleared by the markets that trade the derivative and its underlying securities. It also introduces several risk-free positions (accounts) that accrue at persistently non-zero spreads with respect to each other and the risk-free rate. In the case of derivatives with counterparty default risk \cite{BK13} introduces an approach referred to as semi-replication, which through the choice of cost components in the money account results in derivative prices that carry arbitrage opportunities in the form of holding portfolio of counterparty's bonds versus a derivative position with it.\\  
\indent This paper derives no-arbitrage expressions for default-risky derivative contracts with and without collateral, avoiding these inconsistencies.
\end{abstract}

\newpage
	\thispagestyle{empty}
	\tableofcontents

\numberwithin{equation}{section}
\section{Introduction}
\noindent Derivative transactions today come with an attached collateral account and recognition for the  inherent counterparty default risk. The post-crisis paradigm for derivatives pricing dictates that collateralized with cash trades should be discounted using the rate payable on the cash collateral, while a bank's own cost of funding should be used for discounting non-collateralized trades.\\

\noindent Despite the seemingly bifurcated approach to discounting, the two follow a single principle - in both cases the discounting rate is a funding cost rate for the dealer bank (just different funding needs). Moreover, discounting derivatives with the funding cost has been the case for a long time and before the crisis (so the post-crisis paradigm is not really new). For many years (indeed decades) Libor was considered to be the cost at which dealer banks would fund themselves unsecured. Over the years, using this unsecured funding rate as the rate that was plugged in for the risk-free rate in the no-arbitrage pricing formulas for derivatives has given Libor the title of a risk-free rate. This, of course, was more of a misnomer and belief than an economic reality\footnote{This misnomer is in fact so prevalent, that the industry and regulatory initiatives for replacing Libor is referred to sometimes as a search for an "alternative risk-free rate".}. Nevertheless, as a result of discounting with this single funding cost, dealers would land on the same price, giving the impression that derivatives were priced according to a no-arbitrage pricing approach which upheld the law of one price, and that trades are discounted with a risk-free rate. By implication then one would be forced to state that the inter-dealer trades were happening in a complete and transparent no-arbitrage market. \\

\noindent Libor, however, was still a cost of unsecured funding for the dealer banks, although it was more of a consensus rate set by the Libor panel of dealers, than a benchmark of market traded securities/rates. This consensus single rate implied that all dealer banks have the same credit quality and cost of funding, were they to fund themselves in capital markets. \\

\noindent The crisis, of course, forced fundamental principles of finance back into practice, whereby the funding cost of each dealer is specific to its balance sheet structure (mix of assets and the capital structure) and quality (earning power of assets). This and the lack of a market mechanism in the inter-dealer market that would guarantee a counterparty on either end of a trade meant that transactions in this market are now akin to lending or borrowing and carry default risk and associated costs.\\

\noindent Although arguments around the inclusion of these costs into derivatives pricing are still ongoing (see for example \cite{risk} and \cite{riskhw}), and a fundamental  approach (or an approach from fundamentals) is still missing, the pricing approach in \cite{pit}, \cite{BK} and \cite{BK13} have become an accepted point of view by major derivatives dealers. \\

\noindent We examine origins of this approach by following the three papers. In Section 2 we examine the treatment of the money account in a general setting. We conclude that no-arbitrage pricing neither provides foundations for, nor supports the structures assigned to the money account, and that in general each structural component of the money account should be a risk-free account accruing at a risk-free rate not to contradict the conditions for no-arbitrage. Section 3 demonstrates that the observations on money accounts apply to the case of default-risky derivatives and, by doing so, derives pricing partial differential equations for default-risky derivatives similar to that of a default-risky bond starting with the setup in \cite{BK13}. As expected, the bilateral nature of default for derivatives and the derivatives-specific recovery rates emerge as the only difference between pricing a default-risky bond and a default-risky derivative. Section 4 elaborates on these two features and argues that for derivatives to go pari passu with unsecured senior bonds, derivatives have to be priced as fully or partially unsecured liabilities, as opposed to fully collateralized trades. Section 4 also argues that the collateral account cannot be part of the dynamic variables in the replication portfolio or part of the money account with a non-risk-free rate of accrual. Instead, Section 4 introduces collateral as part of an exogenous recovery process, treating it through the collateralization level as a parameter in the valuation formulas for a default-risky derivative. Section 5 concludes with a summary and remarks on the introduction of XVAs within this approach in the future.\\

\newpage
\section{On the Nature of the Money Account}

\noindent \cite{pit} and then \cite{BK}, \cite{BK13} introduced an approach for incorporating costs related to collateral and hedging of counterparty default risk into a  no-arbitrage (or risk-neutral) price of a derivative. The approach advocated basically amounts to prescribing a structure to the money account, by "taking inspiration" from the costs incurred in managing a typical OTC derivative transaction on a bank's balance sheet. No-arbitrage pricing being the only mechanism (so far) for pricing derivatives neither provides foundations for, nor supports this inspiration. Also implied is that what was known as a position in a risk-free asset (usually proxied with a money account or a risk-free bond) can now be looked at as an account on a bank's balance sheet with components designed to cover the cost structure of a bank's derivatives position, with each component returning bank-specific premiums.\\

\noindent To examine this approach, we start from a most general setup by writing the prescription of a structure to the money account as
\begin{equation}\label{mdistr}
\rho M = \sum _{a} M_{a} r_{a}.
\end{equation} 
\noindent Here, $M = \left\lbrace M_{a}\right\rbrace_{a=1}^{n}$ are structural components of the money account accruing at different rates $r_{a}$, and  $\rho$ is the composite rate of return on the overall money account $M$ that follows by construction. Formally, the results in \cite{pit}, \cite{BK}, \cite{BK13} can be obtained by specifying terms on the right hand side of Eqn.\ref{mdistr} following cost structures in these papers (See Appendix A for details.).\\

\noindent So what can be said about the nature of structural components prescribed to the money account, the accrual rates on these money account components, and the overall rate of return on the money account itself.\\

\noindent We start with the standard Black-Scholes-Merton no-arbitrage pricing setup ("BSM") (\cite{M} and \cite{BS}) as our working model. A self-financing arbitrage portfolio $\Pi$ is set up with a position $h_{V}$ in the derivative $V$ and its replicating portfolio. The latter consists of a position $h_{S}$ in a $\delta$ dividend paying underlying security $S$ and a position in a risk-free security, proxied with a money account $M$. Using the general form of the money account distribution in Eqn.\ref{mdistr} instantaneous return on such a portfolio when it is delta-hedged is given by
	\begin{eqnarray}\label{prtfrtrnHT7}
	\nonumber d\Pi^{\tilde{h}} &=& \left[  \mu_{V} V -   (\mu + \delta) \dfrac{\sigma_{V}}{\sigma} V \right] h_{V} dt + dM\\
	\nonumber &=& \left[  \mu_{V} V -   (\mu + \delta) \dfrac{\sigma_{V}}{\sigma} V \right] h_{V}  dt +  \sum _{a} dM_{a} \\
\nonumber	&=& \left[  \mu_{V} V -   (\mu + \delta) \dfrac{\sigma_{V}}{\sigma} V  \right] h_{V} dt +  \sum _{a} M_{a} r_{a} dt\\
	&=& \left[  \mu_{V} V -   (\mu + \delta) \dfrac{\sigma_{V}}{\sigma} V \right] h_{V} dt + \rho M dt.
	\end{eqnarray}
\noindent Here, being delta-hedged has its usual meaning - i.e., special weights $\tilde{h}_{S} = - \dfrac{\sigma_{V}}{\sigma} \dfrac{V}{S} h_{V}$ and $\tilde{h}_{V} = h_{V}$ exist in the market $\left\lbrace V, S, M \right\rbrace$. Notice, that $\tilde{h}_{S} \neq - \dfrac{\partial V}{\partial S} h_{V} \equiv \Delta h_{V}$, since we have not yet assumed that V = V(t, S(t)). We have only assumed that both the derivative and the underlying security price the same risk factor in all states of the world.\\

\noindent With a zero initial investment (\textbf{\textit{z.i.i. constraint}} henceforth), one has 
\begin{equation}\label{zinv0}
\Pi^{\tilde{h}} (t) = 0 \Rightarrow M = - \tilde{h}_{S} S - \tilde{h}_V  V = \left( \dfrac{\sigma_{V}}{\sigma}  - 1 \right) h_{V} V := \sum_{a} M_{a} ,
\end{equation}
leading to the following form of Eqn.\ref{prtfrtrnHT7}
	\begin{equation}\label{prtfrtrnGEN1}
	d\Pi^{\tilde{h}} = \left[ \mu_{V} V -   (\mu + \delta) \dfrac{\sigma_{V}}{\sigma} V + \left( \dfrac{\sigma_{V}}{\sigma} - 1 \right) V \rho \right] h_{V} dt.
	\end{equation}
\noindent Requiring the no-arbitrage prices to exist 
\begin{equation}\label{no_arb}
 \mathbb{P} \left[  \Pi^{\tilde{h}} (t + dt) \right] = \mathbb{P} \left[ \Pi^{\tilde{h}} (t) \mid_{\Pi^{\tilde{h}} (t) = 0} + d\Pi^{\tilde{h}} \right] = \mathbb{P} \left [ d\Pi^{\tilde{h}} = 0 \right] = 1,
\end{equation}
is equivalent to setting the square brackets in Eqn. \ref{prtfrtrnGEN1} to zero. This yields the following risk-return condition for the traded in this market derivatives and their underlying securities
	\begin{equation}\label{BSMrho1a}
\dfrac{\mu_{V} - \rho}{\sigma_{V}} = \dfrac{(\mu + \delta)- \rho}{\sigma}. 
	\end{equation}
\noindent Eqn.\ref{BSMrho1a} is the condition for no-arbitrage prices to exist in the $\left\lbrace V, S, M \right\rbrace $ market. It is a more fundamental relationship than its BSM PDE representation. To arrive to a BSM PDE representation of Eqn.\ref{BSMrho1a} and its Feynman-Ka\v{c} solution, one still has to make two essential assumptions - 1) $V = V \left( t, S(t)\right) $ and 2) $V \left( T, S(T)\right) = Contractual \; Payoff \equiv \Phi \left( T, S(T) \right) $. After this,  using Ito's Lemma produces a PDE form of the Eqn.\ref{BSMrho1a} 
	\begin{equation}\label{BSMrho2a} 
	\frac{\partial V}{\partial t} + \left( \rho - \delta \right)  S \frac{\partial V}{\partial S} + \dfrac{1}{2} \sigma^{2} S^{2} \frac{\partial^{2} V}{\partial S^{2}} - \rho V = 0.
	\end{equation}
\noindent The only difference here with the classical BSM equation is the composite rate of return $\rho = \sum _{a} \dfrac{M_{a}}{M} r_{a}$ for the money account. \\

\noindent In this sense, the no-arbitrage setup itself does not fix the rate of accrual for the money account. It is rather the risk-free nature of the money account that dictates that its rate of accrual should be the risk-free rate (i.e., $\rho = r$) to avoid arbitrage between risk-free securities (accounts), that accrue at premium differentials to each other. \\ 
 
\noindent In XVA literature money accounts are sometimes constructed bottom-up. First, its components are defined and accrual rates are assigned according to some costs structure arguments (or inspirations). Then, the sum of these component money accounts is defined as the money account - $\sum_{a} M_{a} : = M$. It should be reminded, however, that the cost structure is usually outside of the replication framework in the market, and for the z.i.i. constraint for a delta-hedged portfolio $\Pi^{\tilde{h}}$ to hold
\begin{equation}\label{zinva}
\Pi^{\tilde{h}} = h_{V} V -\Delta h_{V} S + M = 0, \; i.e., \; M = -h_{V} V + \Delta h_{V} S = \left( \Delta S - V \right) h_{V},
\end{equation}
\noindent the $\left\lbrace V, S, M \right\rbrace $ market has to clear
\begin{equation}
-\dfrac{h_{V} V}{M} + \dfrac{\Delta h_{V} S}{M} = 1 = \sum_{a} w_{a}, \; where \; w_{a} = \dfrac{M_{a}}{M}.
\end{equation}
\noindent In \cite{pit}, for example, this is maintained by adding and subtracting the collateral account for collateralized derivatives (with $h_{V} = -1$)
\begin{equation}
\dfrac{(V - C + C)}{M} - \dfrac{\Delta S}{M} = \dfrac{ (V - C)}{M} -\dfrac{C}{M} -  \dfrac{\Delta S}{M} = \sum_{a} w_{a} = 1. 
\end{equation} 
\noindent Although this can be looked at as a mathematical identity, it implies replication of the difference between a derivative $V$ and its collateral account $C$ with a position in the underlying and a risk-free asset 
\begin{equation}\label{pitclear}
V - C = \Delta S + (M - C) = \Delta S + M^{'},
\end{equation}
which does not follow from any fundamental statement about replicability of a contingent claim in this market (collateral prices no risk factor).\\

\noindent More importantly, the z.i.i. constraint Eqn.\ref{zinva}, which implied by the feasible replicability relationships in a given market, defines the size of the money account $M$ and it does not prescribe a structure $M_{a}$ to it (i.e., z.i.i. constraint holds for the total of the money account). \\

\noindent To study Eqn.\ref{BSMrho1a} further, let's rewrite the instantaneous return Eqn.\ref{prtfrtrnGEN1} for the delta-hedged self-financing arbitrage portfolio $\Pi^{\tilde{h}}$ by combining the general form for the money account structure Eqn.\ref{mdistr} and the z.i.i. constraint Eqn.\ref{zinv0} 
\begin{equation}\label{prtfrtrnGEN3}
d\Pi^{\tilde{h}}  =  \left[ \dfrac{\mu_{V} - r}{\sigma_{V}} - \dfrac{\mu + \delta - r}{\sigma} + \dfrac{ \left( \dfrac{\sigma_{V}}{\sigma} -1 \right)  \sum _{a} w_{a} (r_{a} - r)}{\sigma_{V}}  \right] h_{V} V \sigma_{V} dt.
\end{equation}
\noindent Requiring for the arbitrage-free prices to exist (i.e., Eqn.\ref{no_arb} to hold), leads to the following form of the fundamental relation Eqn.\ref{BSMrho1a}
\begin{equation}\label{capmarb}
\dfrac{\mu_{V} - r}{\sigma_{V}} + \dfrac{ \left( \dfrac{\sigma_{V}}{\sigma} -1 \right)  \sum _{a} w_{a} (r_{a} - r)}{\sigma_{V}} = \dfrac{\mu + \delta - r}{\sigma}.
\end{equation}
\noindent The extra premium for the same level of risk $\sigma_{V}$ on the left hand side of Eqn.\ref{capmarb} is due to non-zero premiums $r_{a} - r = \epsilon_{a} \geq 0$ embedded in the (risk-free) components of the money account. It states that extra premium(s) can be earned by holding the derivative instead of a position in the underlying.\footnote{One could argue that a derivative position on a security is always more volatile (more risky) than an outright position in the underlying security and so the price of a derivative should always be greater or equal to the price of the underlying. This would make the derivative a dominant security (\cite{M}), however, the dominance cannot be due to dealer-specific cost structure add-ons to the market clearing derivative prices.}. In other words, one hedges all of the $\sigma_{V}$ in the market for the underlying, get compensated  per unit of derivative position hedged through 
\begin{equation}
\nonumber \mu_{V} =	r + \left( \mu + \delta - r\right) \dfrac{\sigma_{V}}{\sigma} = r +  \left( \mu + \delta - r\right) \left|  \dfrac{\tilde{h}_{S} S}{h_{V} V} \right|_{\tilde{h}_{S}=-\Delta h_{V}} = r +  \left( \mu + \delta - r\right) \left|  \dfrac{\Delta S}{ V} \right| ,
\end{equation}
\noindent and still makes extra premiums on top of a risk-free rate of return, despite the fact that at that point it is a risk free portfolio itself (i.e., with weights $\tilde{h}_{S}=-\Delta h_{V}$).\\

\noindent One can also look at the Eqn.\ref{prtfrtrnGEN3} slightly differently (leading to the same argument). No-arbitrage pricing is effectively a proof that no-arbitrage prices exist in a given market, as opposed to a statement that all securities in that market are traded at no-arbitrage prices. However, if an equilibrium is attainable in the complete market $\left\lbrace V, S, M \right\rbrace $ (i.e., all agents have been able to attain thier optimal risk-return balances), then one should put 
\begin{equation}\label{capmr}
 \dfrac{\mu_{V} - r}{\sigma_{V}} = \dfrac{\mu + \delta - r}{\sigma},
\end{equation}   
\noindent and one could state that markets have cleared at no-arbitrage prices. \\

\noindent The two arguments above imply that for the no-arbitrage prices to clear in the complete market $\left\lbrace V, S, M \right\rbrace$, the second term on the left hand side of Eqn.\ref{capmarb} should be zero with certainty
\begin{equation}\label{prtfrtrnGEN4}
\mathbb{P} [d\Pi^{\tilde{h}} = \left[ \dfrac{ \left( \dfrac{\sigma_{V}}{\sigma} -1 \right)  \sum _{a} w_{a} (r_{a} - r)}{\sigma_{V}}  \right] h_{V} V dt = 0] = 1,
\end{equation}
\noindent leading to
\begin{equation}
r_{a}  - r = 0, \; for\; all\; a\;, and \; \rho = r. 
\end{equation}
\noindent Notice that we have not made an a priori assumption that $\rho = r$ in the derivations of Eqns.\ref{capmarb} - \ref{prtfrtrnGEN4}. \\

\noindent In other words, if one has assumed that $M_{a}$ structural components of the money account are risk-free, then all components $M_{a}$ must accrue at the risk-free rate $r$ to avoid arbitrage - several portfolios presumed risk-free are returning at persistently non-zero premiums with respect to each other. \\

\noindent Although we refer to Eqn.\ref{prtfrtrnGEN3} and Eqn.\ref{capmarb} as arbitrage, the more accurate statement would be that the market $\left\lbrace V, S, M \right\rbrace $ does not clear the equilibrium (no-arbitrage) relation Eqn.\ref{capmarb}
\begin{equation}\label{capmarb1}
\dfrac{\mu_{V} - r}{\sigma_{V}} + \dfrac{ \left( \dfrac{\sigma_{V}}{\sigma} -1 \right)  \sum _{a} w_{a} (r_{a} - r)}{\sigma_{V}} \neq \dfrac{\mu + \delta - r}{\sigma}.
\end{equation}
\noindent Eqn.\ref{capmarb} would point at arbitrage if the term  $\dfrac{ \left( \dfrac{\sigma_{V}}{\sigma} -1 \right)  \sum _{a} w_{a} (r_{a} - r)}{\sigma_{V}}$ on its left hand side was generated by a market-priced security in the $\left\lbrace V, S, M \right\rbrace $ market (as part of the replicating portfolio).\\

\noindent The correct equilibrium (no-arbitrage) expression is Eqn.\ref{capmr}, which states that there is no extra unit of risk-adjusted premium to be received for holding a derivative position, compared to the outright holding of a position in the underlying, if the markets are arbitrage-free (and complete).\\

\noindent This does not prohibit from choosing a benchmark $\rho = r + \epsilon$ to compare the returns from holding a position in the derivative - $\dfrac{\mu_{V} - \rho}{\sigma_{V}}$, or a position in the underlying - $\dfrac{\mu + \delta - \rho}{\sigma}$, and still write
\begin{equation}\label{capmarb2}
\dfrac{\mu_{V} - r - \epsilon}{\sigma_{V}} = \dfrac{\mu + \delta - r - \epsilon}{\sigma}.
\end{equation}
\noindent This simply means the returns over a chosen benchmark $\rho$ are lesser exactly by the amount of $\epsilon = \rho - r$ (investor specific cost above the risk-free rate). The latter has no reflection on the risk-return equilibria in the market  $\left\lbrace V, S, M \right\rbrace $. \\

\noindent To summarize observations above, the money account can have any number of structural components as long as
\begin{enumerate}
	\item[(i)] each component of the money account is a risk-free account, 
	\item[(ii)] each component accrues at a risk-free rate, and
	\item[(iii)] all these risk-free rates are the same.
\end{enumerate}

\noindent  Observations above and the structure of the arbitrage portfolio $\Pi$ also imply that a collateral account cannot be part of the replicating portfolio either as part of the risky positions (the hedge) because it prices no risk factor, or the money account because it's accruing at a non-risk-free rate. \\

\noindent In the next section we show that observations (i) - (iii) also hold for the case of default-risky derivatives, and derive pricing equations for the default-risky derivatives with and without collateral. We show  that the money account in this case is simply a bigger account compared to the case of non-default-risky derivatives pricing of BSM.\\

\noindent We mention in passing, that in the case of no risk-free assets in the market the money account and its components can only be zero-beta portfolios, accruing at the same single rate of return for a zero-beta portfolio (\cite{black}). Observations (i) - (iii) also hold true with respect to zero-beta portfolios. \\

\newpage
 \section{Pricing Derivatives with Counterparty Risk}
\noindent We start from the widely accepted assumption in the practitioner literature that the value of a traded derivative depends on the bi-lateral default risk of the counterparties involved
\begin{equation}\label{bi}
\hat{V} =\hat{V} (t; S(t), J_{A}(t), J_{B}(t) ).
\end{equation} 
\noindent The setup\footnote{$J_{A}$ and $J_{B}$ are indicators of default for counterparty $A$ and $B$ respectively, and take value 1, if the corresponding counterparty is in default and zero otherwise.} of a self-financing arbitrage portfolio $\Pi^{h}$ in this case consists of $h_{\hat{V}}$ quantity of a derivative instrument $\hat{V}$ on a security $S$, with its replicating portfolio consisting of $h_{S}$ quantity of the security, $h_{B}$ quantity of risky bonds $P_{B}$ issued by the counterparty $B$, $h_{A}$ quantity of risky bonds $P_{A}$ issued by counterparty A and an amount $M$ of money account\footnote{Here we are following the setup in the XVA literature that originates from \cite{BK}. Fundamentally, a bilateral default is replicated through a first-to-default CDS on the counterparties, since recovery is triggered by the first default event by either of the parties.}. The value of this portfolio at time $t$ is 
\begin{equation}\label{port}
\Pi^{h} (t) = h_{S} S (t) + h_{\hat{V}} \hat{V} (t) + h_{A} P_{A} (t) + h_{B} P_{B} (t) + M(t),
\end{equation}
\noindent with the instantaneous return given in terms of gain processes as
\begin{equation}\label{dfprtfrtrn0}
d\Pi^{h} = h_{S} dG_{S} +h_{\hat{V}} dG_{\hat{V}} + h_{A} dG_{A} + h_{B} dG_{B}  + dG_{M}.
\end{equation}
\noindent The gain dynamics of the portfolio positions are defined as
\begin{eqnarray}\label{gain}
\nonumber dG_{S} &=& dS + dD_{S} = dS + \delta S dt = (\mu + \delta) S dt+ \sigma S dz \\
\nonumber dG_{\hat{V}} &=& d\hat{V} + 0 = \mu_{\hat{V}} \hat{V} dt + \sigma_{\hat{V}} \hat{V} dz + \Delta \hat{V}_{A} dJ_{A}+ \Delta \hat{V}_{B} dJ_{B} \\
\nonumber dG_{A} &=& dP_{A} + r_{A} P_{A} dt = - (1 - R_{A})P^{-}_{A}dJ_{A} + r_{A} P_{A} dt,\\
\nonumber dG_{B} &=& dP_{B} + r_{B} P_{B} dt = - (1 - R_{B})_{B}dJ_{B} + r_{B} P_{B} dt, \\
dG_{M} &=& 0 + dD_{M} = \rho M dt. 
\end{eqnarray}
\noindent Here, as in the previous section, we have taken $M$ to be a money account with generic components $M_{a}$ and a composite accrual rate $\rho$ as in Eqn.\ref{mdistr}. $R_{A}$ and $R_{B}$ are the recovery rates for the senior unsecured bonds issued by counterparties $A$ and $B$ respectively. It is assumed that the bonds are senior unsecured debentures to discuss the fact that derivatives go pari passu with this type of debt in default. In general, bonds $P_{A}$ and $P_{B}$ can be any debentures issued by the counterparty, since what is being replicated is the default risk, while the recovery levels drive the hedge ratios for full coverage of losses in case of default.\\

\noindent $\Delta \hat{V}_{B}$ and $\Delta \hat{V}_{A}$ are changes in the value of the derivative position due to default by the counterparty $B$ and counterparty $A$, respectively:
\begin{eqnarray}\label{deltaBC1}
\Delta \hat{V}_{B} &=& \hat{v}_B - \hat{V}(t; S, dJ_{A} = 0, dJ_{B} = 0), \\ 
\nonumber \Delta \hat{V}_{A} &=& \hat{v}_{A} - \hat{V}(t; S, dJ_{A} = 0, dJ_{B} = 0), 
\end{eqnarray}
\noindent with $\hat{v}_{A}$ and $\hat{v}_{B}$ the residual values of the risky derivative $\hat{V}$, when the counterparty $A$ or counterparty $B$ is in default, respectively
\begin{eqnarray}\label{vavb1}
\hat{v}_{B} (u) &:=& \hat{V} (t; S, dJ_{A} = 0, dJ_{B} = 1),\\
\nonumber \hat{v}_{A} (u) &:=& \hat{V} (t; S, dJ_{A} = 1, dJ_{B} = 0).
\end{eqnarray}
\noindent Plugging Eqn.\ref{gain} into the instantaneous return equation Eqn.\ref{dfprtfrtrn0} yields to the following
\begin{eqnarray}\label{BK12prtfHT1}
d\Pi^{h} &=& \left[ h_{S} (\mu + \delta) S + h_{\hat{V}} \mu_{\hat{V}} V \right] dt + \left[ h_{B} r_{B} P_{B} + h_{A} r_{A} P_{A} \right] dt + \rho M dt\\
\nonumber &+&  \left[ h_{S} \sigma S + h_{\hat{V}} \sigma_{\hat{V}} V \right] dz \\
\nonumber &+&  (1 - R_{B}) \left[ h_{\hat{V}} \hat{V} \dfrac{\Delta \hat{V}_{B}}{\hat{V}(1 - R_{B})} -  h_{B} P_{B} \right] dJ_{B} \\
\nonumber &+& (1 - R_{A}) \left[ h_{\hat{V}} \hat{V} \dfrac{\Delta \hat{V}_{A}}{\hat{V}(1 - R_{A})} - h_{A} P_{A}  \right] dJ_{A}.
\end{eqnarray}
\noindent Observe now, that $\dfrac{\Delta \hat{V}_{A}}{\hat{V}}$ and $\dfrac{\Delta \hat{V}_{B}}{\hat{V}}$ are negative numbers and are percentage drops in the value of the derivative position due to defaults by $A$ and $B$, respectively. Hence, \textbf{define recovery rates $\chi_{B}$ and $\chi_{A}$ for the derivative position} as
\begin{equation}\label{chiBC}
\nonumber \dfrac{\Delta \hat{V}_{B}}{\hat{V}} = - (1 - \chi_{B}) , \; and \; 
\dfrac{\Delta \hat{V}_{A}}{\hat{V}} = - (1 - \chi_{A}). 
\end{equation}
\noindent We further introduce \textbf{loss ratios $z_{A}$ and $z_{B}$  - the ratio of loss rate from a derivative position to loss rate from a bond position of a counterparty}
\begin{equation}\label{zBC}
\nonumber z_{A} = \dfrac{1 - \chi_{A}}{1-R_{A}},\; and \; 
z_{B} = \dfrac{1 - \chi_{B}}{1-R_{B}}.
\end{equation}
\noindent In other words, we are modeling recovery rates from the derivative position as a percentage of the value of the default-risky derivative prior to default
\begin{eqnarray}
\hat{v}_{A} &=& \chi_{A}  \hat{V}(t; S, dJ_{A} = 0, dJ_{B} = 0), \\ 
\hat{v}_{B} &=& \chi_{B}  \hat{V}(t; S, dJ_{A} = 0, dJ_{B} = 0).
\end{eqnarray}
\noindent Rewriting Eqn.\ref{BK12prtfHT1} with these new notations one arrives to
\begin{eqnarray}\label{BK12prtfHT2}
\nonumber d\Pi^{h} &=& \left[ h_{S} (\mu + \delta) S + h_{\hat{V}} \mu_{\hat{V}} V + h_{B} r_{B} P_{B} + h_{A} r_{A} P_{A} \right] dt + M \rho dt \\
\nonumber &+&  \left[ h_{S} \sigma S + h_{\hat{V}} \sigma_{\hat{V}} V \right] dz \\
\nonumber &+&  (1 - R_{B}) \left[ - h_{\hat{V}} \hat{V} z_{B} -  h_{B} P_{B} \right] dJ_{B} \\
&+& (1 - R_{A}) \left[ - h_{\hat{V}} \hat{V} z_{A} - h_{A} P_{A}  \right] dJ_{A}.
\end{eqnarray}
\noindent If the market $\left\lbrace \hat{V}, S, P_{A}, P_{B}, M  \right\rbrace$ clears, then the \textbf{the z.i.i. constraint }
\begin{equation}\label{zinv1}
\Pi^{h} = h_{S} S + h_{\hat{V}} \hat{V} + h_{B} P_{B} + h_{A} P_{A}  + M = 0, 
\end{equation}
is a constraint on portfolio weights $h$ such that satisfy
\begin{equation}
- h^{'}_{S} S - h^{'}_{\hat{V}} \hat{V} - h^{'}_{B} P_{B} - h^{'}_{A} P_{A} = M.
\end{equation}
\noindent Plugging into Eqn.\ref{BK12prtfHT2} one arrives to the following expression for the instantaneous return on such portfolio
\begin{eqnarray}\label{BK12prtfHT3}
d\Pi^{h^{'}} &=& \left[ h^{'}_{S} (\mu + \delta - \rho) S + h^{'}_{\hat{V}} \mu_{\hat{V}} V \right] dt\\
\nonumber &+& \left[h^{'}_{B} ( r_{B} - \rho )P_{B} + h^{'}_{A} ( r_{A} - \rho ) P_{A} \right] dt \\
\nonumber &+&  \left[ h^{'}_{S} \sigma S + h^{'}_{\hat{V}} \sigma_{\hat{V}} V \right] dz \\
\nonumber &+&  (1 - R_{B}) \left[ - z_{B} h^{'}_{\hat{V}} \hat{V}  -  h^{'}_{B} P_{B} \right] dJ_{B} \\
\nonumber &+& (1 - R_{A}) \left[ - z_{A} h^{'}_{\hat{V}} \hat{V} - h^{'}_{A} P_{A}  \right] dJ_{A}.
\end{eqnarray}
\noindent It is easy to see now, that a portfolio strategy\footnote{We have used the fact that $\hat{V} =\hat{V} (t; S(t), J_{A}(t), J_{B}(t) )$ and so $\dfrac{\sigma_{\hat{V}}}{\sigma} \dfrac{\hat{V}}{S} = \dfrac{\partial \hat{V}}{ \partial S} \equiv \Delta$.} 
\begin{equation}
\tilde{h} = \left\lbrace -\Delta h^{'}_{\hat{V}}, h^{'}_{\hat{V}}, - z_{B} \dfrac{\hat{V} }{P^{-}_{B}} h^{'}_{\hat{V}},  - z_{A} \dfrac{\hat{V} }{P^{-}_{A}} h^{'}_{\hat{V}}, 1 \right\rbrace
\end{equation}
that satisfies the z.i.i. constraint Eqn.\ref{zinv1}
\begin{equation}\label{zinvd}
M = \left( \Delta S - \hat{V} + z_{B}  \hat{V} +  z_{A}  \hat{V} \right) h^{'}_{\hat{V}},
\end{equation}
\noindent  will make instantaneous return on such portfolio equal to
\begin{equation}\label{instcoll1}
d\Pi^{h^{'}} \mid_{h^{'} = \tilde{h} } = \left[(\mu_{\hat{V}} - \rho ) \hat{V} - \Delta (\mu  + \delta - \rho) S  - z_{B} ( r_{B} - \rho) \hat{V} - z_{A} ( r_{A} - \rho) \hat{V} \right] h^{'}_{\hat{V}} dt
\end{equation}
\noindent \textbf{with certainty}. \\

\noindent We rewrite the delta-hedging weights explicitly for a later use
\begin{equation}\label{hedge1}
\tilde{h}_{S} = - \Delta h^{'}_{\hat{V}}, \; and \; \tilde{h}_{\hat{V}} = h^{'}_{\hat{V}};
\end{equation}
\begin{equation}\label{hedge1a}
\tilde{h}_{A} = - z_{A} \dfrac{\hat{V} h^{'}_{\hat{V}} }{P_{A}}  = - \dfrac{(1 - \chi_{A})}{(1-R_{A})} \dfrac{\hat{V} h^{'}_{\hat{V}}}{P_{A}},
\end{equation}
\begin{equation}\label{hedge1b}
\tilde{h}_{B} = - z_{B} \dfrac{\hat{V} h^{'}_{\hat{V}}}{P_{B}}  = - \dfrac{(1 - \chi_{B})} {(1-R_{B})} \dfrac{\hat{V} h^{'}_{\hat{V}}}{P_{B}}.
\end{equation}

\noindent Satisfying the no-arbitrage conditions 
\begin{equation}\label{noarb0}
\mathbb{P} \left[  \Pi^{\tilde{h}} (t + dt) \right] = \mathbb{P} \left[ \Pi^{\tilde{h}} (t) \mid_{\Pi^{\tilde{h}} (t) = 0} + d\Pi^{\tilde{h}} \right] = \mathbb{P} \left [ d\Pi^{\tilde{h}} = 0 \right] = 1
\end{equation}
is now equivalent to setting 
\begin{equation}\label{sharperisky}
(\mu_{\hat{V}} - \rho ) \hat{V} - \Delta (\mu  + \delta - \rho) S - z_{B} ( r_{B} - \rho) \hat{V} - z_{A} ( r_{A} - \rho) \hat{V} = 0.
\end{equation} 
\noindent If one assumes that the market $\left\lbrace \hat{V}, S, P_{A}, P_{B}, M \right\rbrace$ is frictionless (no material liquidity premiums), complete and risk-neutrally priced, then
\begin{eqnarray}
 \mathbb{E} \left[ dJ_{A} \right] &\approx& \lambda_{A}dt \; and \;\mathbb{E} \left[ dJ_{B} \right] \approx \lambda_{B}dt, \; with\\
 r_{A} - r &=& \left( 1 - R_{A} \right) \lambda_{A} \; and \;  r_{B} - r = \left( 1 - R_{B} \right) \lambda_{B},
\end{eqnarray}
\noindent and following the arguments in Section 2 for Eqns.\ref{capmarb} - \ref{prtfrtrnGEN4},  $\rho - r = \sum_{a} w_{a} (r_{a} - r) = 0$. \\

\noindent This leads to the following form of the fundamental no-arbitrage relationship Eqn.\ref{BSMrho1a} for the case of default-risky derivatives\footnote{Notice that $\lambda_{A,B}$ are not the same as in \cite{BK}, as the rate of return on collateral account $r_{C}$ is not the risk-free rate - $r_{C} \neq r$.}
\begin{eqnarray}\label{riskycapm1}
\nonumber	\dfrac{\mu_{\hat{V}} - \left[  \left( 1- \chi_{A} \right) \lambda_{A} + \left( 1- \chi_{B} \right) \lambda_{B} \right]  - r}{\sigma_{V}} &=& \dfrac{(\mu + \delta)- r}{\sigma}, \;or\\
\dfrac{\mu_{V} - r}{\sigma_{V}} &=& \dfrac{(\mu + \delta)- r}{\sigma},
\end{eqnarray}
\noindent where $\mu_{V} = \mu_{\hat{V}} - \left[  \left( 1- \chi_{A} \right) \lambda_{A} + \left( 1- \chi_{B} \right) \lambda_{B} \right]$ is the expected return on an otherwise identical default-risk-free derivative $V(t, S(t))$.\\

\noindent Eqn.\ref{riskycapm1} is a no-arbitrage condition for the market $\left\lbrace \hat{V}, S, P_{A}, P_{B}, M \right\rbrace$ and it states that if default risk between two default-risky counterparties can be replicated through bilateral exchange of bonds that are also traded in the market (mutually shorting bonds), then \textbf{after a market-priced adjustment for the bilateral default risk} there is no extra unit of risk-adjusted premium to be obtained for holding a default-risky derivative position, instead of holding an outright position in the underlying, if the markets are arbitrage-free (and complete).\\

\noindent In contrast, prescribing a structure to the money account $M$, using the anzats Eqn.\ref{mdistr} leads to  
\begin{eqnarray}\label{riskycapm2}
\nonumber	\dfrac{\mu_{\hat{V}} - \left[  \left( 1- \chi_{A} \right) \lambda_{A} + \left( 1- \chi_{B} \right) \lambda_{B} \right]  - r}{\sigma_{V}} &+& \dfrac{ \left[ \left( \dfrac{\sigma_{V}}{\sigma} -1 \right) + \dfrac{1 - \chi_{A}}{1-R_{A}} +\dfrac{1 - \chi_{B}}{1-R_{B}} \right]  \sum _{a} w_{a} (r_{a} - r)}{\sigma_{V}} \\
\nonumber  &\neq& \dfrac{(\mu + \delta)- r}{\sigma}, \;or\\
\nonumber \dfrac{\mu_{V} - r}{\sigma_{V}} &+& \dfrac{ \left[ \left( \dfrac{\sigma_{V}}{\sigma} -1 \right) +\dfrac{1 - \chi_{A}}{1-R_{A}} +\dfrac{1 - \chi_{B}}{1-R_{B}} \right]  \sum _{a} w_{a} (r_{a} - r)}{\sigma_{V}} \\
 &\neq& \dfrac{(\mu + \delta)- r}{\sigma},
\end{eqnarray}
\noindent which does not hold  for the same reasons discussed in Section 2 for the case of no default risk.\\

\noindent The market  $\left\lbrace \hat{V}, S, P_{A}, P_{B}, M \right\rbrace$ simply does not clear no-arbitrage prices of a derivative and its replicating portfolio in a way that Eqn.\ref{riskycapm2} between the risk premia in that market would hold.\\

\noindent Eqn.\ref{riskycapm1} can also be written in a PDE presentation using Ito's Lemma for $\hat{V} \mu_{\hat{V}}$ and $\hat{V} \sigma_{\hat{V}}$ in the standard form with the assumption Eqn.\ref{bi}  
\begin{equation}\label{BSMrisky1}
\mathcal{L}^{\left( r - \delta \right) } \hat{V} - r \hat{V} = \left[ (1 - \chi_{B})  \lambda_{B}  + (1 - \chi_{A})  \lambda_{A} \right]  \hat{V},
\end{equation}
\noindent where, for the brevity of expressions, we have adopted the notation\footnote{Formally, one could still write a PDE form for the expression Eqn.\ref{riskycapm2} \begin{equation*}\label{BSMrisky} 
\mathcal{L}^{\left( \rho - \delta \right) } \hat{V} - \rho \hat{V} = \left[ z_{B}  (r_{B}  - \rho ) + z_{A} (r_{A} - \rho )\right] \hat{V}, \end{equation*} but its solution will not represent a no-arbitrage price of a derivative that is cleared by a complete market.}
\begin{equation}
\mathcal{L}^{\left( x-y \right) } \hat{V} : = \frac{\partial \hat{V}}{\partial t} + \left( x - y \right)  S \frac{\partial \hat{V}}{\partial S} + \dfrac{1}{2} \sigma^{2} S^{2} \frac{\partial^{2} \hat{V}}{\partial S^{2}}.
\end{equation}

\noindent With the appropriate terminal conditions one can write a Feynman-Ka\v{c} solution for Eqn.\ref{BSMrisky1} as follows
\begin{eqnarray}\label{FK1}
\hat{V} (t,S(t), J_{A}(t), J_{B}(t))&=& \mathbb{E}^{Q} \left[ Z_{r}(t, T) \Phi (T, S(T)) \right] \\
\nonumber &-& \mathbb{E}^{Q} \left[ Z_{r}(t, u)  \left( 1 - \chi_{B} \right)  \lambda_{B}(u)  \hat{V} (T - (t + u), S(u)) du \right]\\
\nonumber &-& \mathbb{E}^{Q} \left[ Z_{r}(t, u)  \left( 1 - \chi_{A} \right) \lambda_{A}(u) \hat{V} (T - (t + u), S(u)) du \right],
\end{eqnarray}
\noindent with the simplified notation $Z_{r}(t, T) =  e^{- \int^{T}_{t} r (\tau)d\tau}$. \\

\noindent Eqn.\ref{BSMrisky1} can also be re-written as
\begin{equation}\label{BSMrisky2a}
\mathcal{L}^{\left( r - \delta \right)} \hat{V} - \left[  r + (1 - \chi_{B})  \lambda_{B}  + (1 - \chi_{A})  \lambda_{A} \right] \hat{V} = 0,
\end{equation}
\noindent resulting in a different form of the Feynman-Ka\v{c} solution
\begin{equation}\label{FK1a}
\hat{V} (t,S(t), J_{A}(t), J_{B}(t)) = \mathbb{E}^{Q} \left[  e^{- \int^{T}_{t} \left[  r(\tau) + (1 - \chi_{B})  \lambda_{B}(\tau)  + (1 - \chi_{A})  \lambda_{A}(\tau) \right] d\tau} \Phi (T, S(T)) \right]. 
\end{equation}

\noindent The following observations are in order.
\begin{itemize}
	\item [\textbf{(A).}] Eqn.\ref{BSMrisky1} and Eqn.\ref{BSMrisky2a} that price default risky bonds (see, for example \cite{DS}) are also the equations that price default risky derivatives. Notice, that only contractual cashflows enter the expressions.
	\noindent Differences with pricing a risky bond come only from
	\begin{enumerate}
		\item[-] the bi-lateral nature of the default event\footnote{See also \cite{DH} and the references therein for the bi-lateral nature of default risk in derivatives.}, and
		\item[-] the recovery rate for a derivative.
	\end{enumerate}

\end{itemize}
\begin{itemize}
	\item [\textbf{(B).}] The money account in the case of default-risky derivatives is simply a bigger account compared to the case of non-default-risky derivatives pricing of BSM
	\begin{eqnarray}
	\nonumber	M = \left( \Delta S - \hat{V} + z_{B}  \hat{V} +  z_{A}  \hat{V} \right) h^{'}_{\hat{V}} &=& \left( \Delta S - \hat{V} \right) h^{'}_{\hat{V}} +\left(z_{B}  \hat{V} +  z_{A}  \hat{V} \right) h^{'}_{\hat{V}} \\
	&&:= M_{BSM} + M_{Defaultrisk},
	\end{eqnarray}
	 where $M_{BSM} = \left( \Delta S - \hat{V} \right) h^{'}_{\hat{V}}$ is the money account from the BSM pricing of non-default-risky derivatives and $M_{Defaultrisk} = \left(z_{B}  \hat{V} +  z_{A}  \hat{V} \right) h^{'}_{\hat{V}}$ is an additional money account component due to the default risk, both accruing at the risk-free rate
	\begin{equation} 
	r M  =  r M_{BSM} + r M_{Defaultrisk}.
	\end{equation}
	In the language of Section 2 one has a composite accrual rate for the money account $\rho = \dfrac{M_{BSM}}{M} r_{BSM} + \dfrac{M_{Defaultrisk}}{M} r_{Defaultrisk}$, but with $r_{BSM} = r_{Defaultrisk} = r$, so it satisfies $\rho - r = \sum_{a} w_{a} (r_{a} - r) = 0$.
\end{itemize}

\noindent In other words, - \textbf{they are of the same risk-free nature as the money account as a whole, and are accruing at the same risk-free rate}.\\

\noindent In the next section we take a closer look at the two main differences between the pricing PDE Eqn.\ref{BSMrisky1} and its bond counterparts - the bi-lateral nature of default and recovery rates for derivative positions. We also demonstrate how collateral enters the pricing equations as a parameter, consistent with the observations above that collateral cannot be part of the replicating portfolio either as part of the hedging positions because it prices no risk factor, or the money account because it's accruing at a non-risk-free rate.\\

\newpage
\section{Bilateral Default, Recovery Rates for Derivatives and Collateral}

\noindent While the default risk between two trading counterparties is always bilateral in nature, the price of the derivative traded does not have to be. Starting from an assumption that the price of a derivative depends on the bilateral default risk of the counterparties involved (Eqn.\ref{bi}), presumes that in general a derivative (its contractual payoff function) may generate liability cashflows for both counterparties. However, if a) the payoff function (contractual cashflows) of a derivative gives rise to liability cashflows to only one of the counterparties, or b) the liability cashflows of counterparties are separable with certainty (i.e., the payoff function is linear in the two streams of liability cashflows), then Eqn.\ref{BSMrisky1} translates into separate unilateral default-risky equations for the counterparties $A$ and $B$ 
\begin{equation}\label{BSMrisky2uni}
\mathcal{L}^{\left( r - \delta \right) } \hat{V}_{A , B} - r \hat{V}_{A , B} = (1 - \chi_{A , B})  \lambda_{A , B} \hat{V}_{A , B}.
\end{equation}
\noindent Here $\hat{V_{A}}$ and $\hat{V_{B}}$ are the prices for derivatives, the payoff function for which are the liability cashflows due from the counterparty $A$ - $\Phi_{A}(T,S(T))$ and counterparty $B$ - $\Phi_{B}(T,S(T))$, respectively\footnote{The trade can be treated as a structured deal, each component priced as a unilateral default-risky derivative - no need for considering bi-lateral expressions. }.\\

\noindent In general, where separability of liability cashflows cannot be achieved, one needs a more general form of the Eqn.\ref{BSMrisky1}
\begin{equation}\label{BSMrisky3}
\mathcal{L}^{\left( r - \delta \right) } \hat{V} - r \hat{V} = \lambda_{B} \left( \hat{V} - \hat{v}_{B} \right) + \lambda_{A} \left(  \hat{V} - \hat{v}_{A}  \right).
\end{equation}
\noindent Here $\hat{v}_{A}$ and $\hat{v}_{B} $ are the residual values of the (same) derivative contract when counterparty A or counterparty B is in default respectively, as defined in Eqn.\ref{vavb1}. Hence, we rewrite the Feynman-Ka\v{c} solution for Eqn.\ref{BSMrisky3} as
\begin{eqnarray}\label{FK}
\hat{V} (t,S(t), J_{A}(t), J_{B}(t))&=& \mathbb{E}^{Q} \left[  Z_{r}(t, T) \Phi (T, S(T)) \right] \\
\nonumber &-& \mathbb{E}^{Q} \left[ \int^{T}_{t} Z_{r}(t, u)  \lambda_{B}(u) \left( \hat{V} (u) - \hat{v}_{B} (u) \right)     du \right]\\
\nonumber &-& \mathbb{E}^{Q} \left[ \int^{T}_{t} Z_{r}(t, u) \lambda_{A}(u) \left( \hat{V} (u) - \hat{v}_{A} (u) \right)  du \right].
\end{eqnarray}
\noindent Estimation of $\hat{v}_{A}$ and $\hat{v}_{B}$ depends on the applicability and existence of one of the two  mechanisms - \textbf{replacement, or recovery}. \\

\noindent In a replacement paradigm the following two cases are of interest. \\

\noindent \underline{Case 1.} Assume there is a market mechanism for providing counterparty replacement (both for defaulted and solvent counterparties) and all market participants pay the same price for a given contractual stream of cashflows (similar to the pre-crisis setup).\\

\noindent In this case, the derivative contract $\hat{V} (u) = \hat{V} (T - (t + u), S(u))$ at any intermediate time $u \in [t, T]$ can be replaced by an identical one with any other counterparty during the life of the derivative 
	\begin{equation}
	\hat{v}_{A} (u) = \hat{v}_{B} (u) =  \hat{V} (u) = \hat{V} (T - (t + u), S(u)).
	\end{equation}
\noindent The default-risk terms in the Feynman-Ka$\check{c}$ solution Eqn.\ref{FK} vanish, leading to  
\begin{equation}
\hat{V} (t,S(t), J_{A}(t),J_{B}(t)) = \mathbb{E}^{Q} \left[  Z_{r}(t, T) \Phi (T, S(T)) \right]= V(t,S(t)).
\end{equation}

\noindent This outcome is equivalent (at least mechanically) to relaxing the assumption that the price of a derivative that promises a given stream of cashflow depends on the default risk of the counterparties trading it.\\

\noindent The experience of the 2007-2009 financial crisis showed that assuming an a priori existence of a market mechanism for replacement can be a material presumption.\\

\noindent \underline{Case 2.} Assume there is no market mechanism for providing a replacement counterparty, but it is possible to replace the derivative contract with an identical one with another solvent counterparty $C$, and the price at which this replacement is available depends (bilaterally) on the default risk of the parties involved (similar to the post-crisis setup).\\

\noindent If $\hat{V}_{X,Y}$ is the value of a default-risky derivative contract that pays $\Phi(T, S(T))$ at maturity between two solvent counterparties $X$ and $Y$, then one can set 
\begin{eqnarray}
\hat{V} (u) &=& \hat{V} (u; S, dJ_{A} = 0, dJ_{B} = 0):= \hat{V}_{A,B} (u), \: as \;a \;notation;\\
\nonumber \hat{v}_{B} (u) &=& \hat{V} (u; S, dJ_{A} = 0, dJ_{B} = 1) := \hat{V}_{A,C} (u),\;as \; a \; replacement \; when \; B \; is\; in \; default;\\
\nonumber \hat{v}_{A} (u) &=& \hat{V} (u; S, dJ_{A} = 1, dJ_{B} = 0) := \hat{V}_{B,C} (u), \;as \; a \; replacement \; when \; A \; is\; in \; default.
\end{eqnarray}
\noindent With these notations one then has for the Feynman-Ka$\check{c}$ solution Eqn.\ref{FK}
\begin{eqnarray}\label{FK4a}
\hat{V} (t; S(t), J_{A}(t),J_{B}(t))&=& \mathbb{E}^{Q} \left[  Z_{r}(t, T) \Phi (T, S(T)) \right] \\
\nonumber &-& \mathbb{E}^{Q} \left[ \int^{T}_{t} Z_{r}(t, u) \lambda_{B}(u)   \left( \hat{V}_{A,B} (u) - \hat{V}_{A,C} (u) \right)    du \right]\\
\nonumber &-& \mathbb{E}^{Q} \left[ \int^{T}_{t} Z_{r}(t, u) \lambda_{A}(u) \left( \hat{V}_{A,B} (u) - \hat{V}_{B,C} (u) \right)  du \right].
\end{eqnarray}	
\noindent One can also rewrite Eqn.\ref{FK4a} as
\begin{eqnarray}\label{FK4b}
\hat{V} (t; S(t), J_{A}(t),J_{B}(t))&=& \mathbb{E}^{Q} \left[  Z_{r}(t, T) \Phi (T, S(T)) \right] \\
\nonumber &-& \mathbb{E}^{Q} \left[ \int^{T}_{t} Z_{r}(t, u)   \left[\lambda_{A}(u)  +   \lambda_{B}(u) \right] \hat{V}_{A,B} (u)  du \right]\\
\nonumber &-& \mathbb{E}^{Q} \left[ \int^{T}_{t} Z_{r}(t, u)  \left[ \hat{V}_{A,C} (u) \lambda_{B}(u)  + \hat{V}_{B,C} (u) \lambda_{A}(u) \right]  du \right].
\end{eqnarray}
\noindent The last term can be interpreted as the replacement cost of replacing the derivative contract with an identical one with counterparty $C$. Notice, however, that there is no feasible way of knowing the new counterparty $C$ a priori, and it will not be possible to transact at an unknown exit price $\hat{V}_{A,C} (u)$ or $\hat{V}_{B,C} (u)$. Notice also that this would have not been an issue, were we to keep the pre-crisis assumption that the inter-dealer market is made up of counterparties with (approximately) same credit quality\footnote{\cite{DH} estimated that there was barely a 1bp spread that could be attributed to the bi-lateral default in interest rate swap transactions (pre-crisis).}, i.e. $\hat{V}_{A,C} = \hat{V}_{B,C}$. \\

\noindent We argue that when either 
\begin{enumerate}
	\item[(a)] there is no market mechanism that guarantees a counterparty allowing to price derivatives at zero loss - $\hat{v}_{A,B} (u) = \hat{V} (u)$, or
	\item[(b)] the replacement counterparty is not known a priori and the loss amount $\hat{V} (u) - \hat{v}_{A,B} (u)$ is ill-defined,
\end{enumerate}
\noindent estimation of $\hat{v}_{A}$ and $\hat{v}_{B}$ requires a \textbf{change of paradigm from replacement to recovery}, where recovery parameters for the use in pricing formulas are estimated exogenously.\\

\noindent Unfortunately, estimation of the recovery rates for derivative positions can itself be a convoluted one due to netting and offsetting clauses, where the actual exposures at default and their netted position are not readily (or a priori) available (see for example \cite{BM} and \cite{xva}). What goes pari passu with senior unsecured debt in resolution procedures is the netted cashflow on the portfolio of transactions at the time of default. Let $\varphi_{net}$ be the netted expected cashflows between the counterparties, and let's denote it as $v^{+}$ when it is an asset to the counterparty A and as $v^{-}$ when it is an asset to the counterparty B
\begin{eqnarray}
\nonumber  v^{+} &=& max \left[ \varphi_{net} , 0\right], \\
v^{-} &=& min \left[ \varphi_{net} , 0\right].
\end{eqnarray}
\noindent In the recovery paradigm, where derivatives go pari passu with the senior unsecured debt (and there is no collateral), one can put
\begin{eqnarray}\label{vavb}
\nonumber \hat{v}_{A}  &=& R_{A} v^{+}, \; and\\
\hat{v}_{B}  &=& R_{B} v^{-}.
\end{eqnarray}
\noindent The amount $\varphi_{net}$ is the only quantity to which recovery rates $R_{A}$ and $R_{B}$ from the unsecured debt of counterparties can be applied.\\

\noindent The XVA literature largely assumes that unsecured senior debt recovery rates $R_{A}$ and $R_{B}$ are reasonable estimates of choice for the recovery rates $\chi_{A}$ and $\chi_{B}$ for a single derivative instrument, mainly due to the pari passu status of the latter.\\

\noindent On the other hand, from the standpoint of estimating recovery rates from risk-neutral prices (expectation of contractual cash flows) it is tempting to set the recovery rates for derivatives as a percentage of the pre-default value of the default-risky derivative
\begin{eqnarray}\label{vavb0}
\nonumber \hat{v}_{A}  &=& \chi_{A} \hat{V}, \; and\\
\hat{v}_{B}  &=& \chi_{B} \hat{V}.
\end{eqnarray}
\noindent This, of course, recovers the familiar expression of Eqn.\ref{FK1} for a default-risky derivative
\begin{eqnarray}\label{FK1a}
\hat{V} (t;S(t), J_{A}(t),J_{B}(t))&=& \mathbb{E}^{Q} \left[  Z_{r}(t,T) \Phi (T, S(T)) \right] \\
\nonumber &-& \mathbb{E}^{Q} \left[ \int^{T}_{t} Z_{r}(t,u) \left[ \left( 1 - \chi_{B} \right)  \lambda_{B}(u) + \left( 1 - \chi_{A} \right) \lambda_{A}(u) \right]\hat{V} (T - (t + u), S(u)) du \right].
\end{eqnarray}
 
\noindent In general, recovery rates are estimated from unsecured defaults and are counterparty specific functions of the structure and quality of defaulted counterparty's balance sheet - capital structure and residual value of assets\footnote{Strictly speaking, because of the exogenous nature of this process it is difficult to refer to Eqn.\ref{FK1a} as risk-neutral price, unless its second term (or its components) is implied from market prices (see also \cite{brigorisk} and \cite{cont}), analogous to bond markets, where the expected loss premium can be implied by the difference between default-risky and default-risk-free bond prices.}, generally not observable from market prices. Collateral agreements can be put in place to make the recovery levels as predictable as possible. Collateralization does not have to make the recovery rate equal to one. Estimation of recovery rates is exogenous to pricing processes. \\

\noindent For the \textbf{recovery paradigm with collateralization} recovery rate $\chi$ for  a derivative has to be adjusted for collateral to make the estimated recovery rate that of an unsecured senior debt, so that at default the mark-to-market values $\hat{V}$ can go pari passu with the unsecured senior debt.\\

\noindent In the recovery paradigm with collateralization one can estimate $\hat{v}_{A}$ and $\hat{v}_{B}$ as follows. If we assume netted collateral posting and collection, then there is a $C(u)$ amount of collateral available to the solvent party at the time of default. If the position is over-collateralized - $\hat{V}(u) < C(u)$, then recovery is the value of the derivative, otherwise it is the collateral amount plus recovery from the unsecured portion - $C(u) + \chi_{A,B} (\hat{V}(u) - C(u))$. This can be presented in a combined manner as follows\footnote{Note that these are not the recovery expressions widely used in the XVA literature. Here $\hat{v}_{A,B}(\chi_{A,B}, C = 0) = \chi_{A,B} \hat{V}$ as in Eqn.\ref{vavb0}, because $\hat{V}$ as a price is kept positive. It does not carry an asset ("+") or liability ("-") sign implicitly.}
\begin{eqnarray}\label{vavb2}
\nonumber \hat{v}_{A} &=& (\hat{V} - C)^{-} + \chi_{A}(\hat{V} - C )^{+} + C, \\
\hat{v}_{B} &=& (\hat{V} - C)^{-}  + \chi_{B} (\hat{V} - C)^{+} + C.
\end{eqnarray}
\noindent Here $x^{+} = max(x,0)$ and $x^{-} = min(x,0)$.\\

\noindent This leads to \textbf{collateral-adjusted derivatives recovery rates} $\tilde{\chi}_{A}$ and $\tilde{\chi}_{B}$
\begin{eqnarray}\label{chiBC3}
\nonumber 1 - \tilde{\chi}_{A} (C_{A}) &=& - \dfrac{\Delta \hat{V}_{A}}{\hat{V}} = \dfrac{\hat{V} - (\hat{V} - C)^{-} - \chi_{A}(\hat{V} - C)^{+} - C  }{\hat{V}} = \left( 1 - \chi_{A} \right) \dfrac{ \left( \hat{V} - C \right)^{+}}{\hat{V}}\\
& = &\left( 1 - k \right)^{+} \left( 1 - \chi_{A} \right)  ,\\
\nonumber 1 - \tilde{\chi}_{B} (C_{B}) &=& - \dfrac{\Delta \hat{V}_{B}}{\hat{V}} = \dfrac{\hat{V} - (\hat{V} - C)^{+}  - \chi_{B} (\hat{V} - C)^{+} - C }{\hat{V}} = \left( 1 - \chi_{B} \right)  \dfrac{\left( \hat{V} - C \right)^{+}}{\hat{V}} \\
&=& \left( 1 - k \right)^{+} \left( 1 - \chi_{B} \right),
\end{eqnarray}
\noindent with $k = \dfrac{C}{\hat{V}}$ as the level of collateralization for the transaction. This also makes the hedge ratios in Eqn.\ref{hedge1a} and Eqn.\ref{hedge1b} explicitly proportional to the non-collateralized portion of the derivative  
\begin{equation}\label{hedge1acoll}
\tilde{h}^{c}_{A} = - (1 - k)^{+} z_{A} \dfrac{h^{'}_{\hat{V}} \hat{V}}{P_{A}},
\end{equation}
\begin{equation}\label{hedge1bcoll}
\tilde{h}^{c}_{B} = - (1 - k)^{+} z_{B} \dfrac{h^{'}_{\hat{V}} \hat{V}}{P_{B}}.
\end{equation}
\noindent Thus, one arrives to \textbf{the collateral-adjusted form} of the main equation  Eqn.\ref{BSMrisky1} for the price of a collateralized default-risky derivative
\begin{eqnarray}\label{BMSRisky2}
\nonumber \mathcal{L}^{\left( r - \delta \right) } \hat{V} - r \hat{V} &=& \left[ (1 - \tilde{\chi}_{A})  \lambda_{A}  + (1 - \tilde{\chi}_{B})  \lambda_{B} \right]  \hat{V} \\
&=& \left[  \left( 1 - \chi_{A} \right)   \lambda_{A} + \left( 1 - \chi_{B} \right)  \lambda_{B} \right] \left( 1 - k \right)^{+} \hat{V},
\end{eqnarray}
\noindent with the corresponding Feynman-Ka\v{c} solution
\begin{eqnarray}\label{FK2}
\hat{V} (t; S(t), J_{A}(t), J_{B}(t)) &=& \mathbb{E}^{Q} \left[  Z_{r}(t, T) \Phi (T, S(T)) \right]\\
\nonumber &-& \mathbb{E}^{Q} \left[ \int^{T}_{t} Z_{r}(t, u) \left[  \left( 1 - \chi_{B} \right)  \lambda_{B} + \left( 1 - \chi_{A} \right) \lambda_{A} \right] \left( 1 - k \right)^{+} \hat{V}(u) du \right], \; or
\end{eqnarray}
\begin{eqnarray}\label{FK2a}
\hat{V} (t; S(t), J_{A}(t),J_{B}(t))&=& \mathbb{E}^{Q} \left[  Z_{r}(t, T) \Phi (T, S(T)) \right] \\
\nonumber &-& \mathbb{E}^{Q} \left[ \int^{T}_{t} Z_{r}(t, u)  \left[  \left( 1 - \chi_{B} \right)  \lambda_{B} + \left( 1 - \chi_{A} \right) \lambda_{A} \right] \left( \hat{V} (u) - C(u )\right)^{+}  du \right]
\end{eqnarray}
\noindent If the posted collateral is not allowed to be netted (e.g., when collateralization includes initial margins) the collateralization level $k$ acquires a counterparty subscript - $k_{A} = \dfrac{C + I_{A}}{\hat{V}}$ and $k_{B} = \dfrac{C + I_{B}}{\hat{V}}$, where the level of collateralization includes the initial margin collateral amounts $I_{A}$ and $I_{B}$.\\

\noindent With a non-netted initial margin the Feynman-Ka\v{c} expression Eqn.\ref{FK2a} obtains the following form\footnote{\noindent More accurately, though, cash collateral amounts $C(u)$ and $I_{A,B}(u)$ at the intermediate times $u$ are from the previous time interval $u - \delta u$ (See also \cite{pykh}).}
\begin{eqnarray}\label{FK2ac}
\hat{V} (t; S(t), J_{A}(t),J_{B}(t))&=& \mathbb{E}^{Q} \left[  Z_{r}(t, T) \Phi (T, S(T)) \right] \\
\nonumber &-& \mathbb{E}^{Q} \left[ \int^{T}_{t} Z_{r}(t, u)  \left( 1 - \chi_{B} \right)  \lambda_{B} \left( \hat{V} (u) - C(u) - I_{B}(u) \right)^{+}  du \right]\\
\nonumber &-& \mathbb{E}^{Q} \left[ \int^{T}_{t} Z_{r}(t, u)  \left( 1 - \chi_{A} \right) \lambda_{A}  \left( \hat{V} (u) - C(u) - I_{A} (u)\right)^{+}  du \right].
\end{eqnarray}

\noindent In summary, the expression Eqn.\ref{FK2ac} for a no-arbitrage price of a default-risky collateralized derivative implies that:
\begin{itemize}
	\item[-] \textbf{no-arbitrage prices for default-risky derivatives should price the unsecured portion of the default-risky derivative, with unsecured recovery rates}, if they are to be pari passu with the senior unsecured debt,
	\item [-]\textbf{ applying senior unsecured debt recovery rates to pricing of a single derivative is generally not accurate,} as these rates should be applied to the closeout netted exposures at default that go pari passu to senior unsecured debt (an exception could be the simple case where exposures are unilateral),
	\item[-] \textbf{the collateral cannot be part of dynamic variables in the replicating portfolio}, either as part of the risky positions (the hedge) because it prices no risk factor, or the money account because it's accruing at a non-risk-free rate,
	\item[-] \textbf{collateral is part of an exogenously estimated parameter -  the recovery rate}, which is generally dependent on the capital structure and the residual value of the assets, both unobservable to the market.
\end{itemize}

\noindent The last two points on collateral make intuitive sense.\\

\noindent There were no market-priced securities in the replicating portfolio that priced the recovery risk directly in all states of the world. Stated otherwise, the self-financing replicating portfolio was set up in a market that did not have securities that were perfectly correlated with a recovery risk factor. For the same reason it does not help introducing new dynamic risk factors to model the recovery rates $\chi_{A}$ and $\chi_{B}$. This constitutes an incomplete market, causing  semi-replication. Such risks can only be mitigated (collateralization and/or guarantees)  through means outside the market transactions\footnote{Or, perhaps, capitalization with a bank's balance sheet, to complete the market.}, not fully hedged.    \\

\noindent This semi-replication should not be confused with the case in \cite{BK13}. The latter approach cannot be called a semi-replication as it simply is a voluntary under-hedging of the default risk of the issuer of the derivative by choosing different from the full replication weights. In other words,  the market in \cite{BK13} is still complete\footnote{A better terminology probably would be incomplete replication in complete markets, as opposed to semi-replication in incomplete markets.}.\\

\noindent This is easily observed if one rewrites the full replication weights $\tilde{h}$ in Eqn.\ref{hedge1} - \ref{hedge1b} in notations of \cite{BK13}
\begin{eqnarray}
\nonumber h_{A,B} P_{A,B} - R_{A,B} h_{A,B} P_{A,B} &:=& P_{A,B} - P^{D}_{A,B},\; with\; h^{'}_{\hat{V}}  = 1, \; leading \; to\\
\tilde{h}_{S} = - \Delta h^{'}_{\hat{V}}, \; and \; \tilde{h}_{\hat{V}} = h^{'}_{\hat{V}}; \: \: &\Longleftrightarrow& \: \: \tilde{h}_{S} = - \Delta, \; and \; \tilde{h}_{\hat{V}} = 1; \\
\tilde{h}_{A} = - z_{A} \dfrac{\hat{V} h^{'}_{\hat{V}} }{P^{-}_{A}}  = \dfrac{(1 - \chi_{A})\hat{V}}{(1-R_{A})P^{-}_{A}} h^{'}_{\hat{V}}, \: \: &\Longleftrightarrow& \: \: P_{A} - P^{D}_{A} = -\Delta \hat{V}_{A} = \hat{V} - g_{A}; \\
\tilde{h}_{B} = - z_{B} \dfrac{\hat{V} h^{'}_{\hat{V}}}{P^{-}_{B}}  = \dfrac{(1 - \chi_{B})\hat{V}}{(1-R_{B})P^{-}_{B}}h^{'}_{\hat{V}}, \: \: &\Longleftrightarrow& \: \: P_{B} - P^{D}_{B} = -\Delta \hat{V}_{B} = \hat{V} - g_{B}.
\end{eqnarray}
\noindent \cite{BK13} choose $V - g_{A} = Full Hedge Weight - \epsilon$, "taking inspiration from funding considerations". Not choosing the full replication weights $\tilde{h}$ creates an arbitrage opportunity between holding a derivative position of a counterparty against a portfolio of counterparty's bonds.\\

\noindent There is also no reason that would follow from no-arbitrage pricing for the specific choices of the bonds and their recovery rates in \cite{BK13} (zero recovery for one counterparty and subordinated debt for the other). The full replication is achievable with any portfolio of counterparty $A$ bonds. If we are interpreting the recovery rates $R_{A}$ and $R_{B}$ as the recovery percentage per dollar of a senior unsecured exposure, and the derivative positions go pari passu with the unsecured senior debt of the defaulted counterparty (which they do), then one puts $\chi_{A} = R_{A}$ and  $\chi_{B} = R_{B}$ and arrives to hedge ratios with $z_{A} = z_{B} = 1$. If the counterparty defaults are hedged using any other combination of the defaulted counterparty's bonds, then the hedge ratios are adjusted accordingly. For example, for hedging with subordinated debt one has $z_{A,B} < 1$ for the hedge ratios since subordinated debt recovers less than the senior debt to which derivatives are pari passu. We discuss the approach in \cite{BK13} in more detail in Appendix B, as it is used widely for deriving XVA expressions (see, for example \cite{GK15}).\\

\noindent Finally, it is worth noting, that \textbf{collateralization can also play a systemic role}. One could \textbf{redefine closeout rules to limit the entitlement of the derivatives recovery exclusively to the recovery from collateral accounts}, with no further recourse to the assets of the defaulted counterparty. This would decouple derivatives trading from the rest of bank's deposit funded balance sheet (e.g., no ring-fencing would be necessary).\\

\newpage
\section{Closing Remarks}  
\noindent Following the approach in \cite{pit} and \cite{BK13} market participants have generated adjustments to derivatives pricing formulas to reflect funding and hedging costs, collectively referred to as XVAs. In a separate effort \cite{BBR} show that these costs can also be recovered if they are assigned as dividends to the replicating market securities. However, in either approach there seems to be no market that would clear these dividend or cost components as part of no-arbitrage prices. In either case requiring XVAs to be included into a price of a derivative effectively amounts to what's referred to as "donations" in \cite{ADS}.\\

\noindent In this paper we have shown that  XVAs do not originate from no-arbitrage pricing\footnote{Although formally, setting the second and third terms in Eqn.\ref{FK2ac} equal to $CVA + DVA$ recovers some members of the familiar XVA family.} as they are not part of a self-financing replicating portfolio of market traded securities (there is no market that clears the prices with XVAs as dividends).\\

\noindent Recently, there have been notable attempts to bring XVAs into the corporate finance and accounting frameworks (\cite{ADS}, \cite{alb} and \cite{KKVA}). \\

\noindent In a forthcoming paper \cite{ht19} we will discuss these efforts and will formulate an approach for recovering \textbf{XVAs as P$\&$L measures of balance sheet consumption} for a derivative transaction on a bank's balance sheet, with no reference to no-arbitrage (risk-neutral) pricing.\\ 

\noindent We will also argue, that \textbf{adding XVAs as costs to the price of a derivative transforms derivatives from a market traded instrument into a banking instrument} (contractual cash flows discounted by an all-in yield).

\newpage

\newpage
\appendix

\LARGE
\textbf{Appendix}

\normalsize
\section{Money Accounts in $\cite{pit}$ $\&$ $\cite{BK}$}

\noindent We explicitly expand the money accounts in \cite{pit} and \cite{BK}.\\

\noindent To arrive to the results of \cite{pit}
\begin{equation}\label{pitPDE}
\mathcal{L} V - \delta  S \frac{\partial V}{\partial S}  =   - r_{R} S \frac{\partial V}{\partial S} + r_{C} C + r_{F} (V - C)
\end{equation} 
one has to solve the z.i.i. constraint (Eqn.\ref{zinva}) with $h_{\hat{V}} = -1$, and add and subtract the collateral account
\begin{equation}\label{zinvpit}
M = - \Delta  S + \left( V - C \right)  + C.
\end{equation}
\noindent The addition and subtraction of the collateral account generates the money account components $M_{R}$, $M_{F}$ and $M_{C}$ which are then assigned accrual rates $r_{a}, a = C, F, R$, motivated by the following cost structure:
\begin{description}
	\item[$M_{R} = - \Delta S$] amount of the underlying security borrowed at the repo rate $r_{R}$, with dividend income of  $\delta$; 
	\begin{equation*}
	dM_{R} = (r_{R} - \delta) M_{R} dt;
	\end{equation*}
	\item[$M_{F} = V - C$] amount to be borrowed/lent unsecured from the treasury desk for collateral, which accrues at the funding rate of $r_{F}$
	\begin{equation*}
	dM_{F} = r_{F} M_{F} dt;
	\end{equation*}
	\item [$M_{C} = C$] the collateral account that accrues at the collateral rate of $r_{C}$;
	\begin{equation*}
	dM_{F} = r_{C} M_{C} dt.
	\end{equation*}
\end{description}
\noindent For the money account in \cite{pit} the standard no-arbitrage conditions in Eqn.\ref{no_arb} lead to
\begin{equation}\label{sharpecof1}
\nonumber \mu_{V} -   (\mu + \delta) \dfrac{\sigma_{V}}{\sigma} =  \dfrac{1}{V} \sum_{a = C,F,R} M_{a} r_{a},
\end{equation}
\noindent which means that one needs to put
\begin{eqnarray}\label{muza}
\nonumber \dfrac{1}{V} \sum_{a} M_{a} r_{a} + \dfrac{\sigma_{V}}{\sigma} r &\stackrel{?}{=}& r, \; or\\
\nonumber  \dfrac{1}{V} \sum_{a} M_{a} r_{a} &\stackrel{?}{=}& r \left( 1 - \dfrac{\sigma_{V}}{\sigma} \right) = r \dfrac{V - \Delta S}{V}, \; or \\
\sum_{a=C,F,R} \dfrac{M_{a}}{V - \Delta S} r_{a} &\stackrel{?}{=}& r. 
\end{eqnarray}	
\noindent This means (the z.i.i. constraint Eqn.\ref{zinvpit} holds)
\begin{equation}
\sum_{a=C,F,R} \dfrac{M_{a}}{\sum_{a=C,F,R} M_{a}} = \sum_{a=C,F,R} w_{a} =1,
\end{equation}
and consequently for Eqn.\ref{muza} 
\begin{equation}\label{muza2}
\sum_{a=C,F,R} w_{a} r_{a} \stackrel{?}{=} r , \; or \;\sum_{a=C,F,R} w_{a} \left(  r_{a} - r \right) \stackrel{?}{=} 0.
\end{equation}
\noindent Eqn.\ref{muza2} states that the "portfolio of funding accounts" that gives the money account a structure in \cite{pit} has to be a risk-free (or zero beta ) portfolio.\\

\noindent The arguments above apply to the case of \cite{BK} with the following definitions for the money account components.
\begin{itemize}
	\item Split the funding component $M_{F}$ of the money account into two pieces to account for any surplus or shortfall cash held by the seller after the own bonds have been purchased:
	\begin{description}
		\item [$\left( - \hat{V} - \Delta \hat{V}_{B} \right)^{+}$] surplus cash held by the seller after the own bonds have been purchased accruing at risk-free rate $r$;
		\begin{equation*}
		dM_{F}^{+} = r M_{F}^{+} dt; 
		\end{equation*} 
		\item [$\left( - \hat{V} - \Delta \hat{V}_{B} \right)^{-}$] shortfall that needs to be funded through borrowing, at the financing rate of $r_{F}$
		\begin{equation*}
		dM_{F}^{-} = r_{F} M_{F}^{-} dt; 
		\end{equation*}
	\end{description}
	\item $M_{R} = - \Delta S$ an account for the underlying security borrowed, accruing at the repo rate $q$, provides a dividend income of  $\delta$;
	\begin{equation*}
	dM_{R} = (r_{R} - \delta) M_{R} dt;
	\end{equation*}
	\item $M_{C}$ an account for the proceeds of shorting the counterparty bond through a repurchase agreement at rate $r$ (it is assumed that the haircut in this repo is zero, so that the repo rate for the counterparty bond can be replaced with a risk-free rate)
	\begin{equation*}
	dM_{C} = r M_{C} dt.
	\end{equation*}
\end{itemize}
\noindent With these notations equations Eqn.\ref{muza} - Eqn.\ref{muza2} follow for the case of \cite{BK}.\\

\section{The Case of Semi-Replication in \cite{BK13}}
\noindent We discuss the semi-replication approach introduced in \cite{BK13}, since this approach is used widely for deriving KVA and MVA expressions (as in \cite{GK15}). The approach cannot be called a semi-replication as it simply amounts to under-hedging the default risk of the "issuer" of the derivative by choice. \\

\noindent We will apply the no-arbitrage pricing approach in the main body of the paper to \cite{BK13}, to show that the semi-replication in this paper allows arbitrage.\\

\subsection{Semi-Replication - a Misnomer}
\noindent \cite{BK13} set up a "hedge portfolio" $\Pi$ (not the same as $\Pi^{h}$ above) as
\begin{eqnarray}\label{BK1}
\Pi(t) = q(t) \cdot h(t) = h_{S} S + h_{A} P_{A} + h_{B} P_{B} + M - C 
\end{eqnarray}
with a strategy that $V + \Pi = 0$.\footnote{"...except, possibly, at issuer default" -note by the authors.}. Then  $\Psi := V + \Pi $ is the arbitrage portfolio $\Pi^{h}$ above with $h_{V} = 1$. \\

\noindent \cite{BK13} introduce a money account distribution as $M = M_{S} + M_{B}$, where $M_{S}$ and $M_{B}$ are assumed to be "financing" the $S$ and $P_{B}$ positions respectively. The latter interpretation provides the justification for extra constraints 
\begin{eqnarray}\label{BK2}
\nonumber h_{B} P_{B} + M_{B} &=& 0\\
h_{S} S + M_{S} &=& 0.
\end{eqnarray}
With Eqn.\ref{BK2} the z.i.i. constraint for $\Psi$ looks as follows
\begin{equation}\label{BK4}
\Psi = V + \Pi = V + \underbrace{h_{S} S + M_{S}}_{ = 0} + h_{A} P_{A} + \underbrace{h_{B} P_{B}  + M_{B}}_{=0} - C = V + h_{A} P_{A} - C = 0.
\end{equation}
Using Eqn.\ref{BK2}, the dynamics of the money accounts $M_{S}$ and $M_{B}$ are  assigned rates of return along lines of the financing arguments, with $q$ and $q_{B}$  the repo rates for financing the $S$ and $P_{B}$ positions:
\begin{eqnarray}\label{BK3}
\nonumber dM_{S} &=& r M_{S} dt = -rh_{S}Sdt \longrightarrow - \left( q - \delta \right) h_{S} S dt;\\
dM_{B} &=& r M_{B} dt = - r h_{B} P_{B} dt  \longrightarrow - q_{B} h_{B} P_{B}dt.
\end{eqnarray}

\noindent \cite{BK13} then write the instantaneous return for the self-financing portfolio $\Psi$ as
\begin{equation}\label{BK5}
d\Psi = dV + d\Pi = dV + h_{S} dS + h_{A} dP_{A} + h_{B} dP_{B}  + dM_{S} + dM_{B} - dC.
\end{equation}
Plugging in the price dynamics into Eqn.\ref{BK5} and using the notations from \cite{BK13} for $h_{A,B} P_{A} - R_{A} h_{A,B} P_{A} := P_{A,B} - P^{D}_{A,B}$ leads to
\begin{eqnarray}\label{BK7}
\nonumber d\Psi &=& dV + h_{S} dS + \left[r_{A}P_{A} + h_{B} \left( r_{B} - q_{B} \right) P_{B} + \left( \delta - q  \right) h_{S} S - r_{C}C \right] dt \\
&+& \left[P^{D}_{A} - P_{A} \right] dJ_{A} + \left[P^{D}_{B} - P_{B} \right] dJ_{B}.
\end{eqnarray}
Now, recalling Eqn.\ref{BK4} and choosing  $R_{B} = 0 \Rightarrow P^{D}_{B} = 0 $ leads to the following expression for $d\Psi$\footnote{Notice, that at this point there is no particular reason for choosing $R_{B} =0$.}
\begin{eqnarray}\label{BK8}
\nonumber d\Psi = dV + d\Pi &=& \left[\mu_{V} V + h_{S} \mu S + r_{A}P_{A} + h_{B} \left( r_{B} - q_{B} \right) P_{B} + \left( \delta - q  \right) h_{S} S - r_{C}C \right] dt \\
\nonumber &+& \left[ \sigma_{V} V + h_{S} \sigma S \right] dz\\
&+& \left[ P^{D}_{A} - C + g_{A} \right] dJ_{A} + \left[  \Delta V_{B} - h_{B} P_{B} \right] dJ_{B}. 
\end{eqnarray}
where we have used $\Delta V_{A} = g_{A} - V$.\\

\noindent One can of course require that all the terms in Eqn.\ref{BK8} be set equal to zero to avoid arbitrage, however, we will follow the paper by setting
\begin{eqnarray}\label{deltah1}
h_{S} &=& - \dfrac{\sigma_{V}}{\sigma} \dfrac{V}{S} = - \Delta,\\
h_{B} &=& \dfrac{\Delta V_{B} }{P_{B}}, \; and \\ 
h_{A} &=& P^{D}_{A} - C + g_{A} = \epsilon \neq 0.
\end{eqnarray}
With these new notations, for the delta-hedged portfolio $\Psi = V + \Pi$ one can write
\begin{eqnarray}\label{BK9a}
\nonumber d\Psi = dV + d\Pi &=& \left[\mu_{V} V - \Delta \mu S + r_{A}P_{A} + \left( r_{B} - q_{B} \right) \Delta V_{B} - \left( \delta - q  \right) \Delta S - r_{C}C \right] dt \\
&+& \left[ P^{D}_{A} - C + g_{A} \right] dJ_{A}. 
\end{eqnarray}
Using the expression for $\mu_{V} V$, adopting the notations $r_{B} - q_{B} = \lambda_{B}$\footnote{Note that $\lambda$ here is different from the one in previous sections, it is defined with respect to $q_{B}$.}, $s_{C} = r_{C} - r$, $\epsilon = P^{D}_{A} - C + g_{A}$ and using the remaining z.i.i constraint Eqn.\ref{BK4} one arrives to the following main expression of \cite{BK13} 
\begin{eqnarray}\label{BK9b}
\nonumber d\Psi = dV + d\Pi &=& \left[\mathcal{L}^{q-\delta} V - s_{C}C - \left(r + \lambda_{A} + \lambda_{B} \right) V + ( g_{A} - \epsilon ) \lambda_{A}  + g_{B} \lambda_{B}   \right] dt \\
&+& \epsilon dJ_{A}.
\end{eqnarray}

\noindent \cite{BK13} state that "We assume that the issuer wants the strategy described above to evolve in a self-financed fashion while he is alive", and that "This implies that the issuer requires the total drift term of $dV + d\Pi$ to be zero."\\

\noindent This produces a PDE in \cite{BK13} from which XVA expressions in the literature are derived and interpreted by others (e.g., \cite{GK15}).\\

\noindent Let's notice now that the no-arbitrage conditions for the self-financing portfolio $\Psi = V + \Pi$ with $\Psi (t) = 0$ would imply that arbitrage exists even if one sets the drift term in Eqn.\ref{BK9b} to zero, since the probability of default for the counterparty $A$ is non-zero - $Prob[dJ_{A} = 1] > 0$, and there is always an $\epsilon$ such that
\begin{eqnarray}
\mathbb{P} \left[ \Psi (t + dt) > 0 \right] = \mathbb{P} \left[ \epsilon dJ_{A} > 0 \right] &>& 0, \; and\\
\mathbb{P} \left[ \Psi (t + dt) \geq 0 \right] = \mathbb{P} \left[ \epsilon dJ_{A} \geq 0 \right] &=& 1.
\end{eqnarray}

\noindent Moreover, due to the "abridged" version of the z.i.i. constraint Eqn.\ref{BK4}
\begin{displaymath}
\Bigg \lbrace \begin{array}{rr}
\Psi = \hat{V} + P^{-}_{A} - C = 0 \\ g_{A} + P_{D} - C = \epsilon. 
\end{array}  \Longrightarrow  \hat{V} - g_{A} + P^{-}_{A} - P_{D} = - \epsilon.
\end{displaymath}
\noindent one can write
\begin{equation}\label{semi9}
\left[ g_{A} - \hat{V} + P^{D}_{A} - P^{-}_{A} \right] dJ_{A} = \epsilon dJ_{A} \neq 0.
\end{equation}
Eqn.\ref{semi9} means that there is an arbitrage opportunity in holding the derivative position vs. the positions in bonds, or vise-versa.\\

\noindent The reason for this is, of course, clear - it is due to the choice of hedge ratios Eqn.\ref{deltah1} , instead of the full hedge ratios $\tilde{h}_{A,B}$ in Eqn.\ref{hedge1a} and Eqn.\ref{hedge1b}. We can rewrite them as
\begin{eqnarray}\label{deltah2}
\tilde{h}_{B} \left( P^{-}_{B} - P^{D}_{B}  \right) -  \Delta \hat{V}_{B} h_{\hat{V}} =  \tilde{h}_{B} \left( P^{-}_{B} - P^{D}_{B}  \right) +  \left( \hat{V}_{B} - g_{B} \right) h_{\hat{V}}  &=& 0, \; and\\
\tilde{h}_{A} \left( P^{-}_{A} - P^{D}_{A}  \right) -  \Delta \hat{V}_{A} h_{\hat{V}}  =  \tilde{h}_{A} \left( P^{-}_{A} - P^{D}_{A}  \right) +  \left( \hat{V}_{A} - g_{A} \right) h_{\hat{V}} &=& 0.
\end{eqnarray}
\noindent to show, that the semi-replication in \cite{BK13} simply generates a shift of $\epsilon$ with respect to the full replication hedge ratio for the counterparty $A$  
\begin{equation}\label{semi8}
\hat{V} - g_{A} = - \left( P^{-}_{A} - P_{D} \right)  - \epsilon = Full \; Replication \; Weight - \epsilon.
\end{equation}
To summarize:
\begin{itemize}
	\item the $\epsilon$ shift off of the full replication ratio for the counterparty $A$ does not come from any no-arbitrage constraints, 
	\item it is simply (voluntary) under-hedging and not a semi-replication.
\end{itemize}

\noindent Furthermore, there is no reason for the specific choice of the bonds in \cite{BK13}.

\subsection{Semi-replication - Choice of Replicating Bonds}

\noindent There is no reason for the specific choices of the bonds and their recoveries in \cite{BK13}, the full replication is achievable with any portfolio of counterparty $A$ bonds.\\

\noindent To see this we will approach it from a more general setting. Assume the counterparty $A$ above has issued $n$ bonds $P_{A,i}^{-}$ with different seniorities - different recovery rates $R_{A,i}$. One can write for these bonds
\begin{equation}\label{semi1}
dG_{A,i} = dP^{-}_{A,i} + r_{A,i} P^{-}_{A,i} dt     = - (1 - R_{A,i})P^{-}_{A,i}dJ_{A} + r_{A,i} P^{-}_{A,i} dt.
\end{equation} 
\noindent This transforms the last term of Eqn.\ref{BK12prtfHT1} into
\begin{equation}\label{semi2}
\left[ h_{\hat{V}} \Delta \hat{V}_{A} - \sum_{i=1}^{n} h_{A,i} (1 - R_{A,i}) P^{-}_{A,i}  \right] dJ_{A} = \left[ h_{\hat{V}} \Delta \hat{V}_{A} + \left( P^{D}_{A} - P^{-}_{A}\right)   \right] dJ_{A}.
\end{equation}
\noindent Here the notations are generalization of the ones used in \cite{BK13}
\begin{equation}\label{semi3}
-\sum_{i=1}^{n} h_{A,i} P^{-}_{A,i}  + \sum_{i=1}^{n} h_{A,i} R_{A,i} P^{-}_{A,i} = P^{D}_{A} - P^{-}_{A}.
\end{equation}
\noindent Setting also $h_{\hat{V}} = 1$ and $\Delta \hat{V}_{A} = g_{A} - \hat{V}$ as in \cite{BK13} one can write
\begin{equation}\label{semi4}
\left[ \Delta \hat{V}_{A} + \left( P^{D}_{A} - P^{-}_{A}\right)   \right] dJ_{A} = \left[ g_{A} - \hat{V} + P^{D}_{A} - P^{-}_{A} \right] dJ_{A}.
\end{equation}

\noindent It is easy to see that weights $h_{A,i}$ exist such that the expression in square brackets of Eqn.\ref{semi4} can be set to zero
\begin{equation}\label{semi5}
g_{A} - \hat{V} + P^{D}_{A} - P^{-}_{A} = 0, \,\, or \,\, \hat{V} - g_{A} = - \left(  P^{D}_{A} - P^{-}_{A} \right), 
\end{equation}
i.e. the loss in value of the derivative due to the default by counterparty $A$ is restored by the recovery from a short position in the portfolio of bonds issued by the counterparty $A$.\\

\noindent Notice, that it is not the individual weights in the portfolio of bonds (consequently, not the seniority or other characteristics of each of the issue) that matter, but rather the total amount of the debt holdings
\begin{eqnarray}\label{semi6}
\nonumber -\sum_{i=1}^{n} h_{A,i} P^{-}_{A,i}  + \sum_{i=1}^{n} h_{A,i} R_{A,i} P^{-}_{A,i} &=&  - h_{A} \sum_{i=1}^{n} v_{A,i} P^{-}_{A,i}  + h_{A} \sum_{i=1}^{n} v_{A,i} R_{A,i} P^{-}_{A,i} \\
&=& h_{A} \left( P^{D}_{A} - P^{-}_{A}\right). 
\end{eqnarray}
\noindent Here $h_{A} = \sum_{i=1}^{n} h_{A,i}$ and $v_{A,i} = \dfrac{h_{A,i}}{h_{A}}$.

\newpage

\end{document}